\documentclass[%
aps,
prl,
superscriptaddress,
tightenlines,
showpacs,showkeys,
a4paper,
longbibliography,
reprint,
notitlepage
]{revtex4-1}
\usepackage[english]{babel}
\usepackage{amssymb,amsmath,stmaryrd,array}

\usepackage{graphicx}



\begin{document}
\DeclareGraphicsExtensions{.eps,.png,.pdf}
\title{%
Effects of polarization azimuth in dynamics of
electrically assisted light-induced gliding of 
nematic liquid-crystal easy axis
}

\author{A.V.~Dubtsov}
\affiliation{%
Moscow State University of Instrument Engineering and Computer Science,
 Stromynka 20, 107846 Moscow, Russia}

\author{S.V.~Pasechnik}
\affiliation{%
Moscow State University of Instrument Engineering and Computer Science,
 Stromynka 20, 107846 Moscow, Russia}
\affiliation{%
Hong Kong University of Science and Technology, 
Clear Water Bay, Kowloon, Hong Kong}

\author{Alexei~D.~Kiselev}
\email[Email address: ]{kiselev@iop.kiev.ua}
\affiliation{%
 Institute of Physics of National Academy of Sciences of Ukraine, 
prospekt Nauki 46, 03028 Kiev, Ukraine}
\affiliation{%
Hong Kong University of Science and Technology, 
Clear Water Bay, Kowloon, Hong Kong}

\author{D.V.~Shmeliova}
\affiliation{%
Moscow State University of Instrument Engineering and Computer Science,
 Stromynka 20, 107846 Moscow, Russia}

\author{V.G.~Chigrinov}
\email[Email address: ]{eechigr@ust.hk}
\affiliation{%
Hong Kong University of Science and Technology,
Clear Water Bay, Kowloon,
Hong Kong}

\date{\today}

\begin{abstract}
We experimentally study 
the reorientation dynamics of the nematic liquid crystal easy axis 
at photoaligned azo-dye films
under the combined action of in-plane electric field and 
reorienting UV light linearly polarized at varying polarization
azimuth, $\varphi_p$.
In contrast to the case where
the light polarization vector is parallel to the initial
easy axis and $\varphi_p=0$,
at $\varphi_p\ne 0$,
the pronounced purely photoinduced reorientation
is observed outside the interelectrode gaps.
In the regions between electrodes
with non-zero electric field,
it is found that
the dynamics of reorientation
slows down with $\varphi_p$
and the sense of easy axis rotation
is independent of the sign of $\varphi_p$.
\end{abstract}

 \maketitle

\section{Introduction}
\label{sec:intro}

It is well known that
external electric, magnetic or light
fields may produce deformations
of liquid crystal (LC) orientational structures 
initially stabilized by anisotropic 
boundary surfaces (substrates). 
There is a variety of 
related Fr\'{e}edericksz-type effects
that are at the heart of
operation of the vast majority of 
modern liquid crystal devices~\cite{Yang:bk:2006,Chigr:1999}.

Among the key factors 
that have a profound effect on behavior of 
the field induced orientational transitions
are the boundary conditions at the substrates. 
These conditions are determined by the anchoring characteristics
such as the anchoring energy strengths and the
\textit{easy axis},
$\mathbf{n}_e$, giving the direction of
preferential orientation of LC molecules at the surface.

In contrast to traditional description of
the Fr\'{e}edericksz-type transitions,
it turned out that 
reorientation processes induced by external fields
may additionally 
involve slow rotation of the easy axis.  
Over the past few decades this slow motion~---~the so-called 
\textit{easy axis gliding}~--~has received much 
attention as a widespread phenomenon
observed in a variety of liquid crystals
on amorphous glass~\cite{Oliveira:pra:1991},
polymer~\cite{Vetter:jjap:1993,Vorflusev:apl:1997,Faetti:epjb:1999,Lazarev:lc:2002,Janossy:pre:2004,
Joly:pre:2004,Faetti:pre:2005,Faetti:lc:2006,Pasechnik:lc:2006,Janossy:pre:2010}   
and solid~\cite{Faetti:epjb:1999,Pasechnik:lc:2006} 
substrates. 

Slow reorientation of the easy axis
also takes place on the photosensitive layers prepared using 
the photoalignment (PA) technique
such as poly-(vinyl)-alcohol (PVA) coatings
with embedded azo-dye molecules~\cite{Vorflusev:apl:1997},
polymer compound poly (vinyl methoxycinnamate)~\cite{Lazarev:lc:2002},
and the azo-dye films~\cite{Pasechnik:lc:2006}.
The PA  technique 
is employed in the manufacturing process of liquid crystal displays 
for fabricating high quality aligning substrates
and uses linearly polarized ultraviolet (LPUV) light
to induce anisotropy of the angular distribution
of molecules in an azo-dye containing photosensitive
film~\cite{Chigrin:bk:2008}.

In PA method, the easy axis is determined by 
the polarization azimuth of the pumping LPUV light,
whereas the azimuthal and polar anchoring
strengths may depend on a number of the governing parameters such as
the wavelength and the irradiation dose~\cite{Kis:pre2:2005}.
So, in a LC cell with the initially irradiated layer, 
subsequent illumination with reorienting light which polarization
differs from the one used to prepare the layer
can trigger the light-induced easy axis gliding. 
Such gliding  may be of considerable interest 
for applications such as LC rewritable devices~\cite{Chig:jjap:2008}.

Recently, it has been found experimentally that
a more complicated effect of electrically
assisted light-induced azimuthal gliding of the
easy axis takes place
on photoaligned azo-dye layers
when irradiation of nematic LC (NLC)
cells with LPUV light is combined with the application 
of ac in-plane electric field~\cite{Pasechnik:lc:2008}.
It was observed that,
at certain combinations of the parameters
such as the amplitude of electric field  $E$, the light
intensity, $I_{UV}$, 
the exposure time, $t_{exp}$, and the doze
of the initial UV irradiation, $D_p$, 
the switching off relaxation considerably slows
down up to few months.
The switching on dynamics of the gliding 
for both the linearly polarized and 
the nonpolarized reorienting light was studied 
in~\cite{Dubtsov:pre:2010}. 
In particular,
the results of the papers~\cite{Pasechnik:lc:2008,Dubtsov:pre:2010} 
demonstrate that the combined effect 
may be used as a tool to tune technical
parameters of LC memory devices. 

So, as compared to the case of 
purely light-induced reorientation of 
the easy axis governed by 
the effect of photoinduced ordering 
in azo-dye layers,
the dynamics of the electrically assisted
light-induced gliding can be additionally influenced by
the electric field, $E$.
Another physically and technologically important parameter
is the \textit{polarization azimuth} 
that characterizes orientation of the polarization vector of 
the LPUV reorienting, $\mathbf{E}_{UV}$. 
In previous studies~\cite{Pasechnik:lc:2008,Dubtsov:pre:2010},
the reorienting light was linearly polarized along
the initial easy axis.

In this paper, 
we present the experimental results 
on the reorientational dynamics of 
the electrically assisted light-induced 
azimuthal gliding of the easy axis measured at
different values of the polarization azimuth.
Our goal is to study 
how the polarization azimuth 
influences the surface mediated
reorientation processes that occur in NLC cells under 
the combined action of LPUV light and in-plane electric field.

\begin{figure}[!tbh]
\centering
\resizebox{90mm}{!}{\includegraphics*{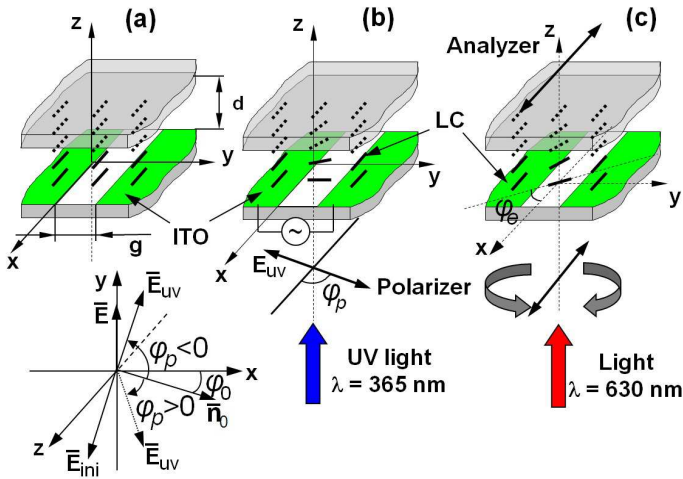}}
\caption{%
Geometry of the experiment: 
(a)~initial state;
(b)~switching on an in-plane electric field  combined with 
LPUV irradiation; 
(c) switching off the electric field and reorienting 
LPUV light to 
measure the easy axis azimuthal angle $\varphi_e$.
}
\label{fig:expt_geom}
\end{figure}

\section{Experiment}
\label{sec:experiment}

In our experiments, liquid crystal (LC) cells ($d=17.4 \pm 0.2\,\mu$m) 
of sandwich like type were assembled 
between two amorphous glass plates. 
The upper glass plate was
covered with a rubbed polyimide film 
to yield the strong planar anchoring conditions.
In Fig.~\ref{fig:expt_geom},
the direction of rubbing
gives the easy axis
parallel to the $x$ axis.
 
A film of the azobenzene sulfuric dye SD1 (Dainippon Ink and
Chemicals)~\cite{Chigrin:bk:2008} 
was deposited onto the bottom substrate
on which transparent indium tin oxide (ITO) electrodes were placed.
The electrodes and the interelectrode stripes
(the gap was about  $g=50~\mu$m) were arranged to be
parallel to the $x$ axis. 

As in~\cite{Pasechnik:lc:2008,Dubtsov:pre:2010},
the azo-dye SD1 layer was initially illuminated by linearly polarized UV
light (LPUV) at the wavelength $\lambda = 365$~nm.
The preliminary irradiation produced the zones of different energy dose exposure 
$D_p=0.27,\,0.55$~J/cm$^2$ characterized by relatively 
weak azimuthal anchoring strength. 
The light propagating along 
the normal to the substrates (the $z$ axis) 
was selected by an interference filter.
Orientation of the polarization vector of UV light,
$\mathbf{E}_{\mathrm{ini}}$, was chosen so as to
align azo-dye molecules at a small angle of 4 degrees to 
the $x$ axis, $\varphi_0\approx 4$~deg 
[see Fig.~\ref{fig:expt_geom}(a)].

The LC cell was filled with the nematic LC mixture E7 (Merck) 
in isotropic phase and then slowly cooled down to 
room temperature. 
Thus we prepared the LC cell
with a weakly twisted planar orientational structure 
where the director at the bottom surface 
$\mathbf{n}_0$ is clockwise rotated  
through the initial twist angle $\varphi_0\approx 4$~deg
which is the angle between $\mathbf{n}_0$ 
and the director at the upper substrate (the $x$ axis).

As is indicated in figure~\ref{fig:expt_geom}(b),
the director field deforms when
the in-plane ac voltage ($U=100$~V, $f=3$~kHz)
is applied to the electrodes. 
In addition to the electric field,
$E=2$~V/$\mu$m,
the cell was irradiated with the reorienting
LPUV light beam 
($I_{UV}=0.26$~mW/cm$^2$ and $\lambda = 365$~nm)
normally impinging onto the bottom substrate.

For this secondary LPUV irradiation,
orientation of the polarization plane is determined by
the \textit{polarization azimuth} 
$\varphi_p$ which is defined as the angle between 
the polarization vector of UV light, $\mathbf{E}_{UV}$, 
and the initial surface director $\mathbf{n}_0$.
As is indicated in Fig.~\ref{fig:expt_geom},
we shall assume that
positive (negative) values of the polarization azimuth,
$\varphi_p>0$ ($\varphi_p<0$),
correspond to clockwise (counterclockwise) rotation 
of the polarizer from $\mathbf{n}_0$ to $\mathbf{E}_{UV}$.

\begin{figure}
\centering
   \resizebox{90mm}{!}{\includegraphics*{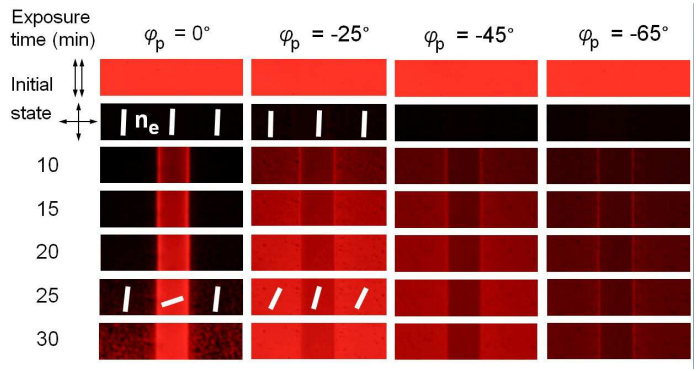}}
\caption{%
Microscopic images of the cell (filter with $\lambda =630$~nm was used)
in crossed polarizers for LPUV irradiation at different
exposure times for various values of the polarization azimuth. 
The interelectrode gap ($g=50\,\mu$m) is indicated.
The ac electric field is $E=2$~V/$\mu$m\,
and the initial irradiation dose is $D_p=0.27$~J/cm$^2$.
}
\label{fig:expt_img}
\end{figure}

Our experimental method has already been described 
in~\cite{Pasechnik:lc:2008,Dubtsov:pre:2010}.
In this method, NLC orientational structures 
were observed via a polarized microscope connected
with a digital camera and a fiber optics spectrometer. 
The rotating polarizer technique was used to measure 
the azimuthal angle
$\varphi_e$ characterizing orientation of the easy axis. 
In order to register microscopic images and 
to measure the value of $\varphi_e$,
the electric field and the reorienting light were switched off 
for about 1 min (see Fig.~\ref{fig:expt_geom}(c)).
This time interval is short enough 
to ensure that orientation of the easy axis remains essentially intact
in the course of measurements.
The measurements were carried out at a temperature of 26$^{\circ}$C.

When the electric field ($E=2$~V/$\mu$m) 
in combination with reorienting LPUV light of the intensity
$I_{UV}=0.26$~mW/cm$^2$ is applied for more than 120 minutes,
we observed the memory effect. 
In this case, after switching off the field and light,
the easy axis did not relax back to its initial state for at least few months.

Figure~\ref{fig:expt_img} shows the microscopic images
obtained at various times of irradiation  
by LPUV reorienting light for four different values of the
polarization azimuth:
$\varphi_p=0$, $-24$, $-45$, $-65$~degrees.
In this case, the initial irradiation dose
is fixed at $D_p=0.27$~J/cm$^2$.  

It can be seen that, 
when the reorienting light is linearly polarized
along the initial surface director $\mathbf{n}_0$ and $\varphi_p=0$,
the brightness of stripes within 
the interelectrode gaps is much higher
as compared to the ones in the region outside the gaps 
where the electric field is negligibly small $E\approx 0$~V/$\mu$m.
So, in  the case of
vanishing polarization azimuth
studied in Refs.~\cite{Pasechnik:lc:2008,Dubtsov:pre:2010},
we arrive at the conclusion that, by contrast
to the electrically assisted light-induced gliding, 
the purely photoinduced reorientation
is almost entirely inhibited.

\begin{figure}
\centering
   \resizebox{90mm}{!}{\includegraphics*{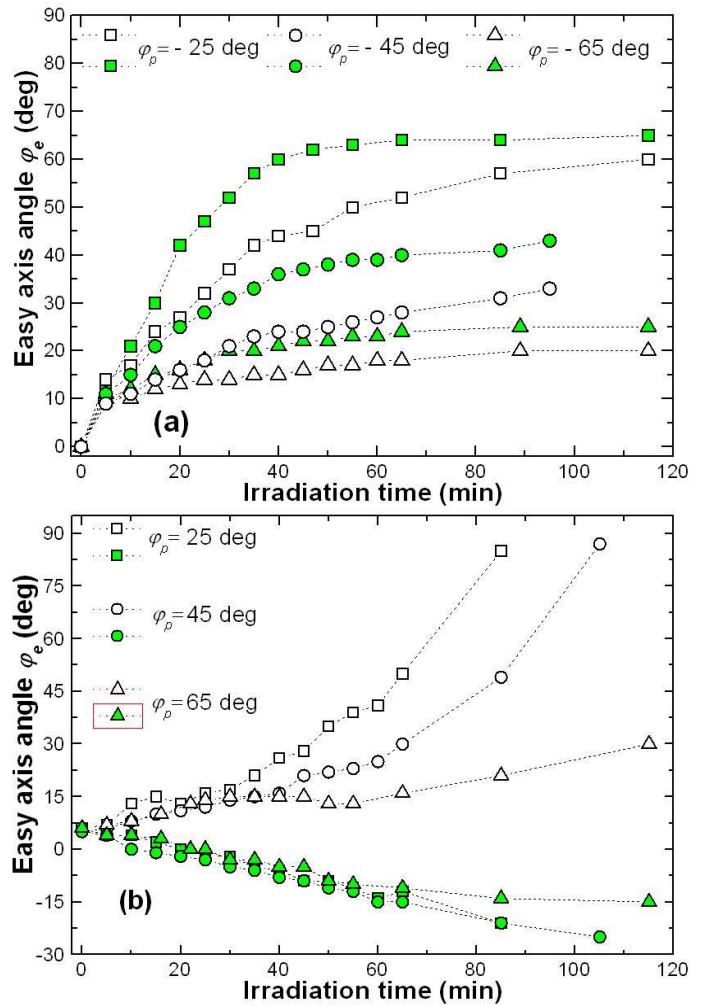}}
\caption{%
Easy axis angle as a function of 
the irradiation time measured
for the reorienting LPUV light
with different 
values of the polarization azimuth $\varphi_p$:
(a)~$\varphi_p<0$ ($D_p=0.27$~J/cm$^2$)
and (b)~$\varphi_p>0$ ($D_p=0.55$~J/cm$^2$). 
Open (solid) circles, squares and diamonds
represent the data measured in the regions within (outside)
the interelectrode gaps
where $E=2$~V/$\mu$m\, ($E\approx 0$~V/$\mu$m).
}
\label{fig:expt_data}
\end{figure}

The latter is no longer the case
for the reorienting light with nonzero polarization azimuth.
Referring to Fig.~\ref{fig:expt_img},
at $\varphi_p\ne 0$, 
light-induced distortions of the surface director in 
the zero-field region located outside the gaps 
are very much more pronounced.
It is also evident from the curves
depicted in Fig.~\ref{fig:expt_data}(a)
representing the irradiation time dependencies of 
the easy axis angle measured
at negative polarization azimuthal angles of 
the reorienting LPUV light. 

In the zero-field curves,
the easy axis angle increases with
the irradiation time
starting from the angle of initial twist,
$\varphi_0$,
and
approaches the photo-steady state
with the photosatured value of the
angle close to $\pi/2+\varphi_p$. 
The curves describing the electrically assisted
reorientation within the interelectrode gaps
lie below the zero-field ones and
reveal analogous behavior.

The data measured at the
polarization azimuths of the opposite sign,
$\varphi_p>0$, (see Fig.~\ref{fig:expt_data}(b))
show that, in the zero-field region,
the light-induced changes of the easy axis angle 
are negative and correspond to the counterclockwise
rotation of the polarizer.
As seen from Fig.~\ref{fig:expt_data}(b),
the dynamics of the easy axis in the presence of the electric field
essentially differs from the one in the regime of purely photoinduced
reorientation.
In the interelectrode gaps,
it turned out that
the electric field prevails thus suppressing 
the tendency for the easy axis to be reoriented
along the normal to the polarization vector of light 
$\mathbf{E}_{UV}$.

\section{Conclusions}
\label{sec:conclusion}

In conclusion, we have experimentally studied 
the effects of polarization azimuth
in the electrically assisted
light-induced azimuthal gliding of the NLC easy axis 
on the photoaligning azo-dye film. 
It is found that,
by contrast to the case where
the polarization vector is oriented
along the initial surface director,
at nonzero polarization angle 
$\varphi_0$,
the purely photoinduced reorientation takes place
outside the interelectrode gaps.
For such field-free regime of reorientation,
the easy axis reorients 
approaching the photosaturation limit
close to the normal to the polarization vector.
These results agree with 
the theoretical predictions
of the diffusion model~\cite{Kis:pre:2009}
describing kinetics of photoinduced ordering in azo-dye films
and the corresponding theoretical analysis will be published 
in a  separate paper.
  
In the regions between electrodes
with non-vanishing electric field,
the dynamics of reorientation slows down with
the polarization azimuth
and, as opposed to the case of 
purely photoinduced reorientation,
the sense of easy axis rotation
for the electrically assisted light-induced gliding 
is found to be independent of the sign of polarization azimuth. 
The above mentioned theory~\cite{Kis:pre:2009}
cannot be directly applied to this case
and the work is in progress on interpreting
these data using 
the phenomenological model formulated in~\cite{Pasechnik:lc:2006,Pasechnik:lc:2008}
to describe the dynamics of the electrically assisted gliding.

\acknowledgments 
This work was partially supported by grants: RF Ministry of Education
and Science: 16.740.11.0324, 14.740.11.0900, 14.740.11.1238;
Development of the Higher Schools Scientific Potential 2.1.1/5873;
HKUST CERG 612310 and 612409.


%

\end{document}